\documentclass[pre,twocolumn,aps,showpacs]{revtex4}
\usepackage{latexsym}
\usepackage{graphicx}
\usepackage[activeacute,english]{babel}                 
                         
\begin{document}
   
\title {Test of the fluctuation theorem for stochastic entropy production in a nonequilibrium steady state} 
           
\author{A. Gomez-Marin\footnote{email: agomezmarin@gmail.com} and I. Pagonabarraga}

\affiliation{Facultat de Fisica, Universitat de Barcelona, Diagonal 647, 08028 Barcelona, Spain} 

\date{\today}

\begin{abstract}
We derive a simple closed analytical expression for the total entropy production 
along a single stochastic trajectory of a Brownian particle diffusing on a periodic potential 
under an external constant force. By numerical simulations we compute  
the probability distribution functions of the entropy and satisfactorily test many of the predictions based on Seifert's integral fluctuation theorem. The results presented for this simple model
clearly illustrate the practical features and implications derived
from such a result of nonequilibrium statistical mechanics.  
\end{abstract}


\maketitle

\section{Introduction}
A great deal of interest has been recently devoted to the development and understanding of formal relationships known as fluctuations theorems.
These expressions generally constrain the probability distribution function of entropy production without relying on how close the system is to thermal equilibrium. 
A number of complementary proposals have been put forward \cite{ev,gc,df,croo,lebo,maes,sst,fara,vzon,jar2}, and a considerable effort has been devoted to clarify the similarities among them. 
Experiments have been carried out to validate and use such expressions to extract useful information in different physical and biological systems \cite{busta,rit}.
A new  fluctuation theorem for dissipative systems has been derived by Seifert \cite{sei}, which
is based on the generalization of the concept of entropy production to the level of a single trajectory. 
 The overall entropy production, namely the sum of the contributions from the environment and the one from the particle along its trajectory, is shown to obey an integral fluctuation relation (and in some cases a detailed one) for any initial conditions, arbitrary driving and over a generic finite time interval.  Moreover, recent experiments have validated this theorem \cite{prlexp1,prlexp2} and the stochastic entropy production has been measured experimentally for the first time \cite{prlexp3}.

In this work, we consider  a Brownian particle moving on a periodic potential under the action of an external constant force. Such system allows us to carry out a detailed study of the total entropy production  and assess carefully the implications of Seifert's fluctuation theorem. After a summary of the main results of Ref. \cite{sei} in Section~\ref{sect:theory}, we particularize them for the system under study in Section~\ref{sect:model}. In Section~\ref{sect:numerics}, we discuss numerical results and compare them with analytic predictions where available. We conclude with a discussion section.

\section{Entropy production along a stochastic trajectory} 
\label{sect:theory}
In Ref. \cite{sei}  the  dynamics of one degree of freedom $x$ such as the position of a particle in a  one dimensional system is studied.  It  evolves according to a Langevin equation,
\begin{equation}
\dot{x} = \mu F(x,\lambda (t)) +\chi,
\label{eq:evol_1}
\end{equation}
where $\chi$ is a Gaussian white noise with autocorrelation $\langle \chi(t)\chi(t')  \rangle = 2 D \delta (t-t')$, and $D$ is the diffusion constant so that in equilibrium $D=T \mu$. We consider $k_B=1$ throughout this work.  
The force $F(x,\lambda (t))=- V'(x)+\phi(x, \lambda(t))$ splits into a
term arising from a conservative  potential, $V(x)$,  plus a time dependent external
source, $\phi(x, \lambda(t))$, whose control parameter is $\lambda (t)$. 
The non-equilibrium Gibbs entropy,
\begin{equation}
S(t) \equiv - \int dxp(x,t) \ln p(x,t) \equiv \langle s(t) \rangle,
\end{equation}
can be interpreted as a mean over single trajectories as specified by the probability $p(x,t)$ of finding  the particle at position  $x$ at time $t$.  Seifert uses the above expression  to propose an entropy   associated to each trajectory of the particle,
\begin{equation}  \label{gibbs}
s(t) = - \ln p(x(t),t).
\end{equation}
The rate of change of entropy of the system is derived from the
Fokker-Planck equation for the probability $p(x,t)$. The rate of heat dissipation in the medium and the increase of entropy in the medium  can be related as $\dot{q}(t)=F(x,\lambda )\dot{x} \equiv T\dot{s}_m$.  Therefore, it is possible to  write down the rate of change of total entropy associated with a single trajectory or realization as
\begin{equation}
\dot{s}_{tot}(t) = \dot{s}_{m}(t) + \dot{s}(t) = -
\left.  \frac{\partial_{t}p(x,t)}{p(x,t)} \right \vert_{x(t)}  +
\left. \frac{j(x,t)}{Dp(x,t)}  \right \vert_{x(t)} \dot{x}, 
\end{equation}
where we have introduced the particle flux, which for the Langevin dynamics we are considering is given by
\begin{equation}
\label{eq:flux_diff}
j(x,t)=\mu F(x,t)-D\frac{\partial p(x,t)}{\partial x}.
\end{equation}

In the spirit of previous fluctuation theorems, which relate the weights of trajectories with their time reversed counterparts,  Seifert introduces the quantity   $R$,
\begin{equation} \label{RR}
R[x(t),\lambda (t);p_{0},p_{1}] \equiv  \ln
\frac{ p[x(t)|x_{0}] p_{0}(x_{0}) }
{ \tilde{p}[\tilde{x}(t)|\tilde{x}_{0}] p_{1}(\tilde{x}_{0}) }, 
\end{equation}
which relates the weight of a trajectory with its time reversed analogue.  The (arbitrary) distributions of initial and final values are $p_{0}(x_{0})$ and $p_{1}(\tilde{x}_{0})$ respectively. 
The tilde refers to the time reversal operation, which associates with a trajectory $x(t)$, a reversed one $\tilde{x}(\tau)=x(t-\tau)$, where $t$ is the duration of the process.
In this way, $\tilde{x}(t)$ retraces the reversed path of the stochastic trajectory $x(t)$, and it starts at $\tilde{x}_0 \equiv \tilde{x}(0) =x(t)$, which corresponds to the final value of the stochastic trajectory $x(t)$.
The system is driven out of equilibrium by an external control parameter that changes with time according to a prescribed protocol $\lambda (t)$, and $p[x(t)|x_{0}]$ and $\tilde{p}[\tilde{x}(t)|\tilde{x}_{0}]$ are the probabilities of the forward and backward paths respectively. 

While $R$ is in principle defined for any normalized distribution $p_{1}(\tilde{x}_0)$, it only becomes the total entropy production (and thus acquires a physical meaning) when $p_1(x_t)=p(x,t)$ is the solution of the Fokker-Planck equation for the initial distribution $p_0(x_0)$. 
Then, the definition (\ref{gibbs}) yields $\Delta s =\ln p_{0}(x_{0}) / p_{1}(x_t)$,  the particle entropy change. The other terms appearing in Eq. (\ref{RR}), related to the path probabilities, correspond to the medium entropy production $\Delta s_m=\ln p[x(t)|x_{0}] / \tilde{p}[\tilde{x}(t)|\tilde{x}_{0}]$, as shown in Ref. \cite{kurch}.
Finally, we can identify $R$ with the total entropy production of the system,
\begin{equation} \label{R+}
R=\Delta s+\Delta s_m=\Delta s_{tot}.
\end{equation}

As we will show in short,  the mathematical identity $\langle e^{-R} \rangle=1$, together with  
Eq. (\ref{R+}), implies the following integral
fluctuation theorem for the total entropy variation along a stochastic
trajectory $\Delta s_{tot}$, 
\begin{equation}
\langle e^{-\Delta s_{tot}} \rangle =1.
\end{equation}
It is absolutely general since it holds at any time along a trajectory generated from generic initial conditions and subject to  arbitrary forces. In the subsequent sections we will analyze the implications and consequences of this relation for a simple example where a detailed analysis of $R$ will be performed.

\section{Derivation of an expression for R}
\label{sect:model}
Let us consider a one dimensional stochastic system in which a particle is
moving in a periodic potential $V(x)$ and under an external  force
$F$ (so that we restrict the general driving to the simplest case of constant driving in time and space, $\phi (\lambda(t),x)=F$) . The corresponding Langevin equation in the overdamped limit is 
\begin{equation} \label{lang}
\dot{x} = -V'(x)+F+\eta(t),
\end{equation}
where the friction coefficient has been absorbed in the time units for simplicity. Then the noise is Gaussian, white, with zero mean and autocorrelation $\langle \eta(t)\eta(t')  \rangle = 2 T \delta (t-t')$.
This system is a simplified version of models analyzed previously in Refs. \cite{hh,rest} to assess some aspects of the fluctuation theorem.

Our aim is to  derive an explicit expression for $R$ in the steady state. To this end, we first split $R$ into two contributions,
\begin{equation} \label{Rsplit}
R= \ln \frac{p[x(t)|x_{0}]}
{\tilde{p}[\tilde{x}(t)|\tilde{x}_{0}]} 
+
\ln \frac{p_{0}(x_{0})}
{p_{1}(\tilde{x}_{0})}. 
\end{equation}
The first term on the rhs involves  the weight of the forward and reversed paths. From the path
integral approach to stochastic systems, it is known that the probability to observe a given realization of the noise is
\begin{equation}
p[ \eta(t) ] \sim exp \left( -\int_{0}^{t}dt' \eta^{2}(t')/4T \right).
\end{equation}
Such a Gaussian dependence can be used, together with the evolution equation (Eq.~(\ref{lang})), to obtain an explicit expression for the transition probability  $p[x(t)|x_{0}]$ in terms of the particle's position. If one takes into account the Jacobian of the transformation, one  arrives at~\cite{kurch}
\begin{equation} \label{entroMed}
\ln \frac{p[x(t)|x_{0}]} {\tilde{p}[\tilde{x}(t)|\tilde{x}_{0}]}=\int_{0}^{t}
(-V'(x)+F)\dot{x}d\tau /T,
\end{equation}
which in the present case reduces to a much simpler form, 
 \begin{equation} \label{path}
\ln \frac{p[x(t)|x_{0}]} {\tilde{p}[\tilde{x}(t)|\tilde{x}_{0}]}
=(F\Delta x - \Delta V)/T, 
\end{equation}
where $\Delta x \equiv  x(t)-x(0) = x_{t}-x_{0}$ and $\Delta V \equiv V(x_{t})-V(x_{0})$.  It is worth mentioning that the final position $x_t$ corresponds to a generic observation time $t$.

The second term on the rhs of Eq. (\ref{Rsplit}) accounts for the ratio of the
probabilities of initial and final positions.  We are interested in the steady state, and thus $p(x)$ follows easily from the corresponding solution of the Fokker-Planck equation.
For a system under an effective potential $U(x)=V(x)-xF$, being $V(x)$ an arbitrary periodic function of period $L$, one gets~\cite{reim}
\begin{equation} \label{px}
p(x)=(\mathcal{N}/T) e^{-U(x)/T}\int_{x}^{x+L}dy
e^{U(y)/T}, 
\end{equation}
where $\mathcal{N}$  ensures proper normalization.
Since the system is in the steady state,  $p_{0}(x)=p_{1}(x)$, which implies
\begin{equation} \label{relationps}
\frac{p_{0}(x_{0})} {p_{1}(\tilde{x}_{0})} =  \frac{p(x_{0})}{p(x_{t})},
\end{equation}
and hence the second term of Eq. (\ref{Rsplit}) can be obtained from Eq.~(\ref{px}).

When inserting Eqs. (\ref{path}), and (\ref{px}) plus (\ref{relationps})  into
(\ref{Rsplit}), a simple expression for $R$ is obtained,
\begin{equation}
\label{Rm}
R=\ln \frac{\int_{x_{t}}^{x_{t}+L}dy e^{U(y)/T}
  }{\int_{x_{0}}^{x_{0}+L}dy e^{U(y)/T} }.
\end{equation}
It shows that $R$ can be computed exactly by integrating the effective potential felt by the particle for every trajectory. It is worth mentioning that this expression is only a function of the initial and final positions, which is a non trivial result.
This fact is of experimental relevance, since one does not need to keep detailed information of the complete trajectory to compute $R$. 
Note also that the  exchange of $x_{0}$ for $x_{t}$ reverses the sign of $R$. We will exploit this important symmetry subsequently.

\begin{figure} 
\begin{center}
  \includegraphics[ angle=270, width=7.5cm]{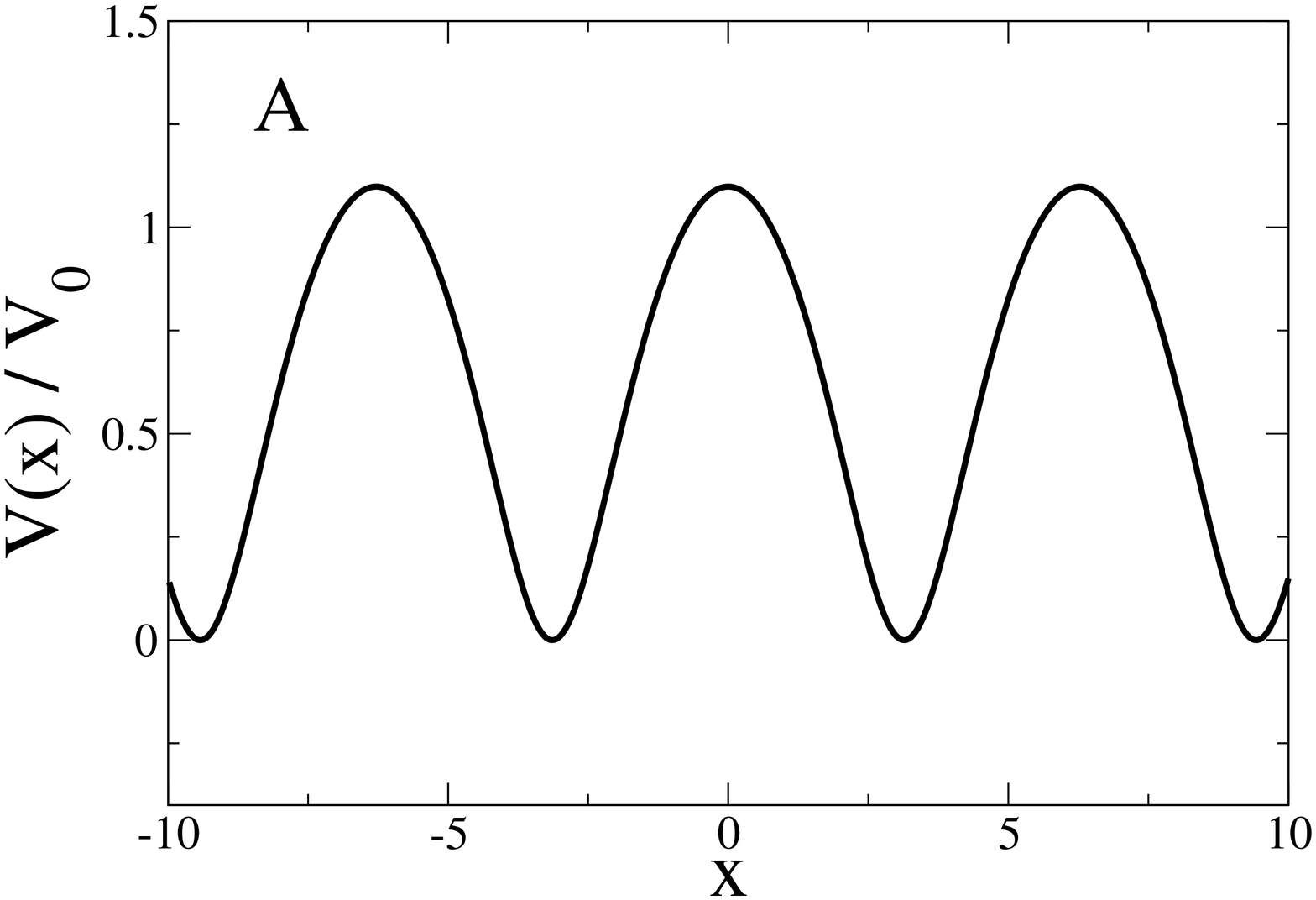}
 \includegraphics[ angle=270, width=7.5cm]{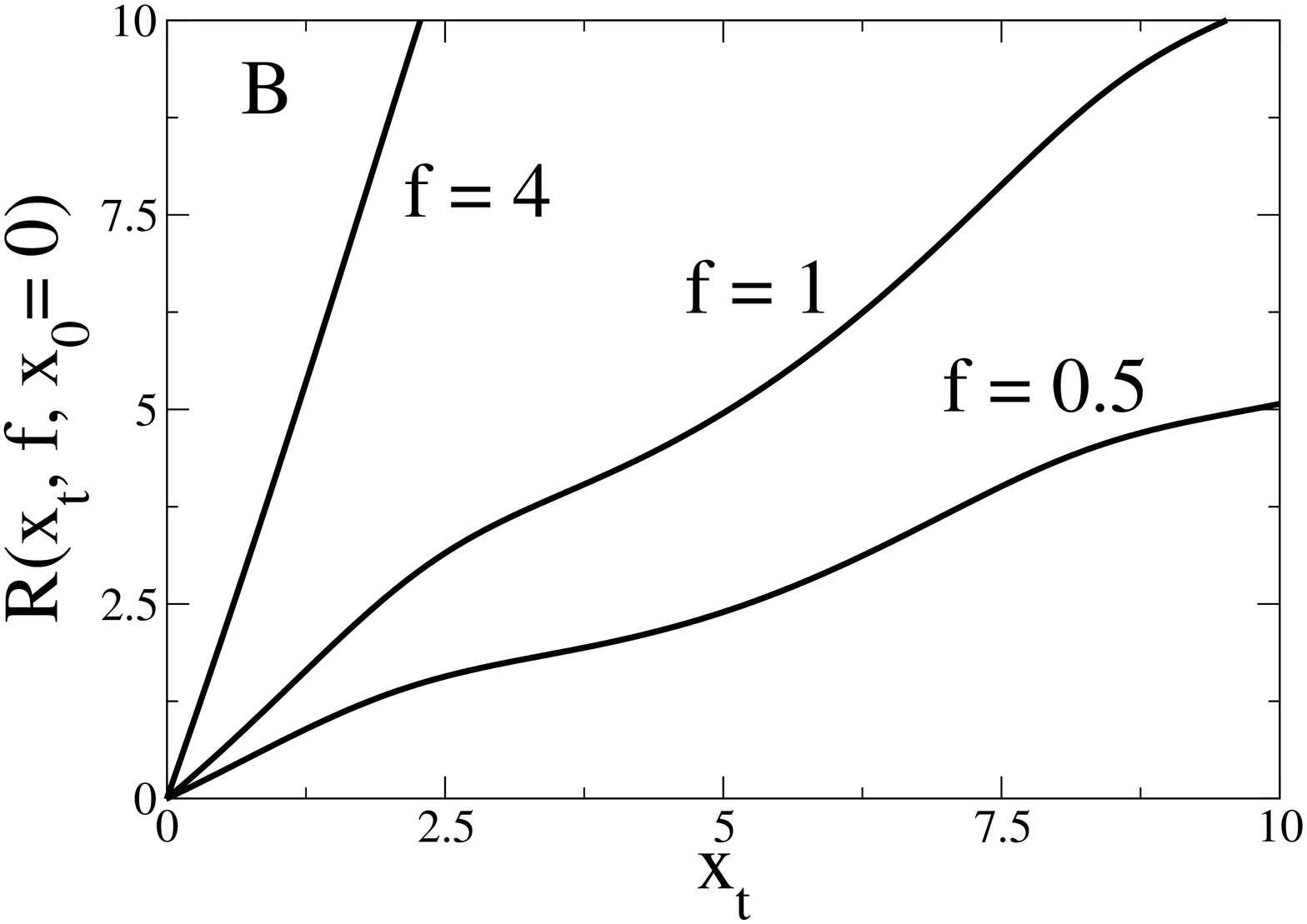}
  \caption{A) Periodic potential defined in Eq. (\ref{pot}) for $L=2\pi$. B) Total entropy variation $R$ obtained from Eq. (\ref{Rmain}) as a function of the final
    position $x_t$ (when the initial position is $x_0 = 0$), for 
    different values of the dimensionless external force parameter $f$.}  
  \label{potential}
\end{center}
\end{figure}

Further progress is feasible if a specific potential is chosen, so that the integrals determining $R$ in the previous equation can be worked out analytically. 
In particular, by picking the potential depicted in Fig. (\ref{potential}.A),
\begin{equation} 
\label{pot}
V(x)= V_{0} \ln (\cos(2 \pi x/L)+2),
\end{equation}
Eq.(\ref{Rm}) reduces to
\begin{equation}
R=\ln \frac{\int_{x_{t}}^{x_{t}+2\pi}dy e^{-y f} \left( \cos(y)+2
  \right)^{v_{0}}   }{\int_{x_{0}}^{x_{0}+2\pi}dy e^{-y f} \left( \cos(y)+2
  \right)^{v_{0}} },  
\end{equation}
where $f=FL/2\pi T$ and $v_{0}=V_{0}/T$. For natural values of the parameter 
$v_{0}$, the integrals can be done analytically, yielding to an explicit expression for $R$ as a function of the initial and final values of the position of the particle for a trajectory taking place during an interval $t$.  Therefore, and without loss of generality, we consider $v_{0}=1$. We then arrive at the simple expression
\begin{equation} 
\label{Rmain}
R=f(x_{t}-x_{0})-\ln \frac{I(x_{t})}{I(x_{0})},
\end{equation}
where
\begin{equation} 
I(x)=f^{2}(\cos(x)+2)-f\sin(x)+2.
\end{equation}
The first term on the rhs of Eq.~(\ref{Rmain}) is directly related to the external work performed on the system by the external force $F$ and then dissipated. The second term has information about the potential $V(x)$ and of the external force too. In Fig. (\ref{potential}.B) we  show the increase of $R$ as a function of the final position along a trajectory, $x_{t}$, when $x_{0}=0$ for different values of the dimensionless driving force $f$. When $f$  is increased, the relative contribution of the periodic potential to $R$ decreases, as can be seen directly from Eq. (\ref{Rmain}). The interesting regime corresponds to the situations in which the  external driving, the periodic potential and the thermal energy have comparable magnitudes. We will concentrate on such regime from now on.

\section{Numerical Results}
\label{sect:numerics}

We perform numerical simulations of Eq. (\ref{lang}) with the potential defined in Eq. (\ref{pot}) 
to characterize the properties of $R$. Such study is required because, even though $R$ depends only on the initial and final positions of the trajectory, the final position is a function of the noise and hence the  probability distribution function of $R$, $p(R)$, has to be obtained.  
Then its properties can be analyzed and the fluctuation theorem tested.

\subsection{Integral fluctuation theorem}
\label{sect:int_th}
\begin{figure} 
\begin{center}
  \includegraphics[ angle=270, width=9cm]{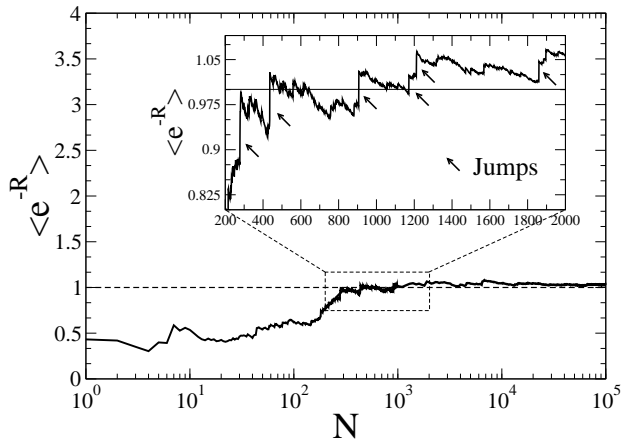}
  \caption{Average of $e^{-R}$ as the number of trajectories $N$ in
    the statistics is
    increased. The parameters are $f=1$ and $t=1$. Convergence is reached quickly. Sudden leaps  are observed as shown in the inset, which is a zoom of the main  figure. Such jumps correspond to the occurrence of rare realizations ($R<0$), which are
    crucial contributions to the average.}
   \label{e-R}
\end{center}
\end{figure}

From the definition of $R$ one can show that
\begin{eqnarray}
\langle e^{-R} \rangle =\sum_{x(t),x_{0}} p[x(t)|x_{0}]p_{0}(x_{0}) e^{-R} =
\nonumber  \\
=\sum_{\tilde{x}(t),\tilde{x}_{0}}
\tilde{p}[\tilde{x}(t)|\tilde{x}_{0}]p_{1}(\tilde{x}_{0}) =1.
\label{eq:general_eR}
\end{eqnarray}
This is a general result, independent of the specific form of the potential and external driving exerted on the particle. 
Using the collection of values of $x_0$ and $x_t$  from the
simulations and inserting them in expression (\ref{Rmain}), we can perform the average
$\langle 
e^{-R} \rangle$, which can be written down as the sum 
\begin{equation}
\langle e^{-R} \rangle = \frac{1}{N} \sum_{i=1}^{N} e^{-R_i},
\end{equation}
where $N$ is the number of particle trajectories that are used to perform the average.
It can also be interpreted as the number of repetitions of the experiment with the same protocol. 
 We display in Fig.~(\ref{e-R}) the average of $e^{-R}$ as a function of the number of trajectories, when the time interval is $t=1$ and the external driving is $f=1$. Around this zone of the space of parameters is where the relation is not trivial at the same time that it is numerically testable.  One can see that the average value predicted  by Eq. (\ref{eq:general_eR}) is rapidly attained. 
The same must hold for any values of $v_0$, $f$ and $t$, as it has been verified for some of them
 (data not shown).
Although  a noisy but progressive  approach to  unity is found as $N$ increases, jumps  are observed, as depicted in the zoom of Fig. (\ref{e-R}).   They  are rare events that  contribute significantly to the average because $R<0$, and hence can be regarded as   transient violations of the second law. We will come back to this point in Fig. (\ref{transient}) when the evolution of $R$ in time for one single trajectory will be discussed.  

\subsection{Probability distribution function of R}  
\begin{figure} 
\begin{center}
  \includegraphics[ angle=270, width=9cm]{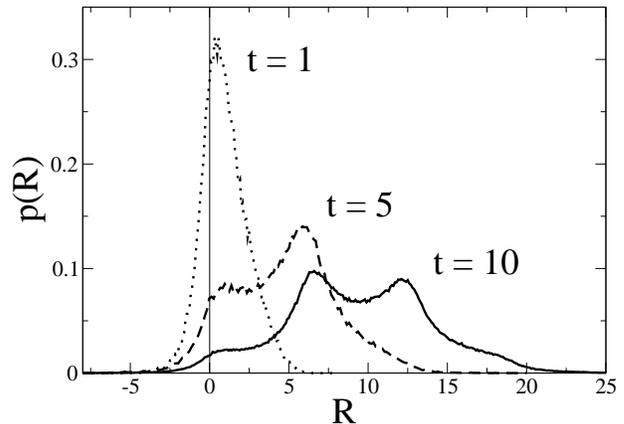}
  \caption{Probability distribution functions of $R$  at different
    times $t$ for an external driving $f=1$.}
  \label{p(R)}
\end{center}
\end{figure}

In order to gain more insight into the properties of $R$ and the implications behind the fluctuation theorem, we 
 analyze the probability distribution of the total entropy production, $p(R)$, as obtained from numerical simulations over many independent trajectories.  Such a quantity can be measured experimentally, and a specific realization of a driven  two level system with time-dependent rates has been already achieved \cite{prlexp2}.
  
Fig. (\ref{p(R)}) exhibits the probability distribution function $p(R)$ at different times, for an external driving $f=1$. The distribution function is peaked and  would reduce to a delta function at $R=0$ at all times  for an equilibrium system. Due to the increase of the mean of $R$ as a function of time, $p(R)$ is shifted towards positive values of $R$ and spreads as time increases. The broadening of the distribution can be attributed to the interaction of the particle with the  underlying periodic potential together with thermal fluctuations. It is interesting to note that $p(R)$ develops a richer structure as time increases. In particular the number of peaks increases because there is a higher chance to explore the periodic potential. Increasing $f$ would lead to a corresponding decrease in this structure in $p(R)$.
For short time intervals, there is a contribution to the distribution that comes from negative values of $R$, in agreement with the jumps described in Fig. (\ref{e-R}). As time increases the contributions to the negative tail decrease, being harder to observe. Precisely, the relation between negative and positive tails is very relevant, as we will discuss in short.

\subsection{Decomposition of $R$}

\begin{figure} 
\begin{center}
  \includegraphics[ angle=270, width=9cm]{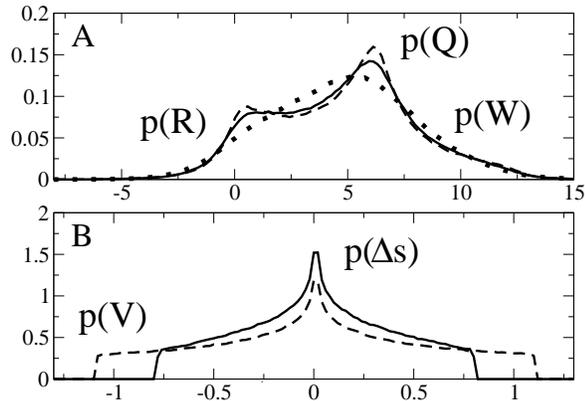}
  \caption{Probability distribution functions (at $t=5$ and $f=1$)  of the several contributions to the total entropy change $R$: A)  Total entropy production $R$ (solid line),
    dissipated heat $Q$ (dashed) and   work  ${\cal W}$ (dotted). B) Internal energy  ${\cal V}$ (dashed) and particle entropy change  $\Delta s$ (solid). Such quantities satisfy $R=Q+\Delta s$ and $Q={\cal W}+{\cal V}$.}
    \label{split}
\end{center} 
\end{figure}

Let's recall that $R$ can be expressed as a sum of two contributions, as written down in Eq.~(\ref{Rsplit}). The first term involves a ratio of conditional probabilities and is inextricably related to the heat $Q$ interchanged with the heat bath \cite{seki}, because $Q/T=\Delta s_{m}= \int_{x(0)}^{x(t)} (\dot{x}-\eta(t'))dx(t')/T $  reduces to Eq. (\ref{path}) when Eq. (\ref{lang}) is used. In turn, $Q$ can be decomposed in two significant contributions; a first term proportional to the work produced by the external force,  ${\cal W}=F\Delta x/T$, and a second one, ${\cal V}=-\Delta V/T$, which corresponds to the change in the internal potential energy of the system. 
The remaining contribution to $R$, the second term on the rhs of Eq. (\ref{Rsplit}), is the entropy change of the particle  $\Delta s$ along the stochastic trajectory.
On physical grounds, it is  fruitful to write $R=Q+\Delta s={\cal W}+{\cal V}+\Delta s$ so  that three different physical sources contribute to the change in the overall entropy of the system. The total entropy
change $R$ has contributions from the medium and from the system itself, namely, the total heat dissipated $Q$ (which comes both from the external forcing and the change of internal energy) and the entropy change associated to the change in the system's internal state, $\Delta s$.
 
As presented in Fig.~(\ref{split}), for the simple model under scrutiny it is possible to study each  probability distribution function of the different contributions separately. Obviously,   although $R=Q+\Delta s$, $p(R) \neq p(Q)+p(\Delta s)$. One can see that  the structure of the probability distribution function $p({\cal W})$, whose contribution is directly the external force, has no internal structure and is simply characterized by a broad, slightly tilted,  peak which moves to higher values of $W$ as time increases.
 $p(Q)$ retains the structure of the peaks, which we attribute to the effect of the periodic potential. See Ref. \cite{agm} for a study of the structure of the probability distribution function of heat exchanged in a simple temperature transducer.  In Fig.~(\ref{split}.B), $p({\cal V})$ is shown. It has a bound support and its shape does not change as time increases. Since the term associated to the external force takes increasing values, it is clear that the  contributions from the periodic potential to the entropy change will have a restricted effect on the overall $p(R)$. Nevertheless, the potential is the responsible for the peaked structure in $p(R)$. The contribution of $\Delta s$ is qualitatively analogous to that of ${\cal V}$.

\subsection{Detailed fluctuation theorem}

\begin{figure} 
\begin{center}
  \includegraphics[ angle=270, width=9cm]{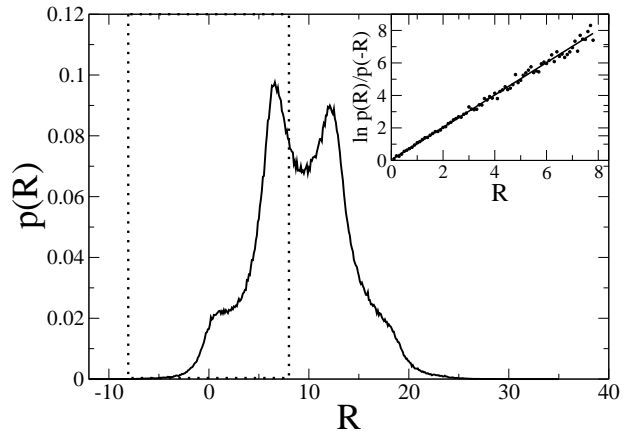}
  \caption{$p(R)$ for $f=1$ at $t=10$. The dotted area shows the part of the probability distribution function used in the inset. The second peak is not included in the range of the inset because the statistics of the corresponding negative values is very poor. The detailed relation $\ln p(R)/p(-R)=R$ is checked satisfactorily.}
   \label{StrongRel}
\end{center}
\end{figure}

A stronger fluctuation relation has also been shown to hold for the probability distribution function of $R$ itself \cite{sei}, 
\begin{equation} \label{lnPR}
\ln \frac{p(R)}{p(-R)} = R.
\end{equation}
It constrains directly the properties of the probability distribution function of  $R$.
This  stringent relation  is closely related to the studies performed in Refs. \cite{gc,ev,croo,jar2}, where similar fluctuation relations are derived for entropy, work and heat in different contexts.
The previous relation follows~\cite{enz}  from the fact that for an arbitrary function $g(R)$, 
\begin{eqnarray} \label{demos}
\langle g(R)e^{-R} \rangle =\sum_{x(t),x_{0}} p[x(t)|x_{0}]p_{0}(x_{0})
\; g(R)e^{-R}= \nonumber \\ 
=\sum_{\tilde{x}(t),\tilde{x}_{0}}
\tilde{p}[\tilde{x}(t)|\tilde{x}_{0}]p_{1}(\tilde{x}_{0}) \;
g(-R[\tilde{x}(t)]) = \langle g(-R) \rangle,
\end{eqnarray}
where the symmetric property of $R$, namely $R[x(t)]=-R[\tilde{x}(t)]$, is used and necessary to prove the above equation. This important condition must be verified  in every case. Here it is satisfied because $R$ reverses sign when interchanging  $x_0$ and $x_t$.  
Then, for the particular case  $g(R) \equiv \delta (R[x(t)]-R)$, one  recovers Eq. (\ref{lnPR}).

In the inset of Fig.~(\ref{StrongRel}) we check the detailed fluctuation theorem by relating the negative and positive contributions to the probability distribution of $R$, leading to an excellent agreement with the theoretical prediction. As we will show in detail, $\langle R \rangle$ increases linearly with time. Therefore, the probability distribution function of $R$ moves towards positive values as time goes by. Then, concerning the verification of the detailed fluctuation theorem, the longer the time of observation, the poorer the statistics at the negative tails, thus limiting the possibility of a meaningful tests. Experimental tests have been performed in Ref. \cite{prlexp1} to verify the detailed fluctuation relation.

\begin{figure} 
\begin{center}
  \includegraphics[ angle=270, width=9cm]{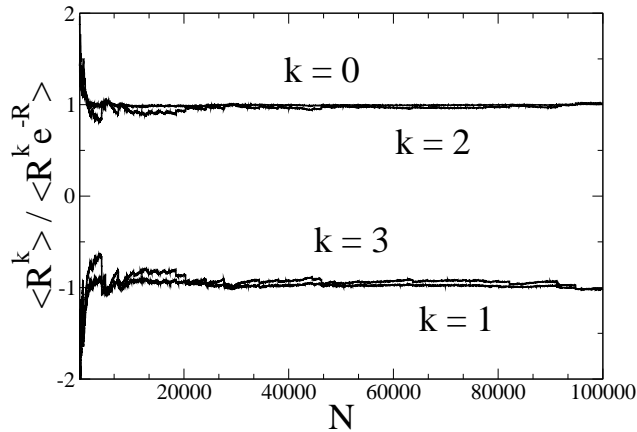}
  \caption{Test of the relation (\ref{mom}) for the zeroth, first,
    second and third moments of $p(R)$ at $t=1$ and $f=1$ as a function of the number of trajectories. }
   \label{moments}
\end{center}
\end{figure}

One can further take profit of Eq. (\ref{demos}) by considering $g(R)=R^k$. A set of relations involving average moments of powers of the total entropy change of the system are found,
\begin{equation} \label{mom}
\langle R^{k}e^{-R} \rangle = (-1)^{k} \langle R^{k}  \rangle.
\end{equation}
In Fig. (\ref{moments}) we display the lower order moments ($k=0,1,2,3$) as a function of the number of independent trajectories, $N$. Theoretical predictions are recovered. The dependence on $N$ shows again the statistical nature of the fluctuation theorem. Note that the behavior is analogous to the one in Fig. (\ref{e-R}). 

\subsection{Initial conditions}

\begin{figure} 
\begin{center}
  \includegraphics[ angle=270, width=9cm]{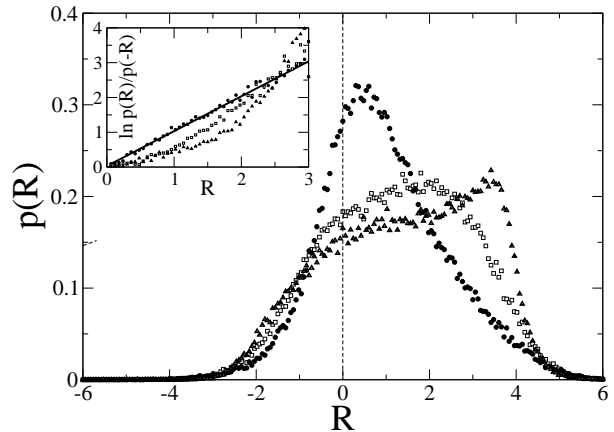}
  \caption{Probability distribution function $p(R)$ for $f=1$ at $t=1$ obtained from different
    ensemble initial conditions: circles correspond to steady state initial
    distribution of positions, triangles to a delta function at
    position zero and squares are obtained from a Gaussian distribution of
    positions. The inset shows that only the probability distribution function
    of $R$ obtained from the steady state ensemble of initial conditions fulfills the
    detailed version of the theorem, Eq. (\ref{lnPR}).}   
  \label{init}
\end{center} 
\end{figure}

Seifert's integral fluctuation theorem is valid for arbitrary initial conditions. However, Eq. (\ref{Rmain}) holds for $R$ only in the steady state. Therefore, when sampling from other different initial conditions, the relations verified so far will not be reproduced anymore. Fig.~(\ref{init}), displays the histograms  at $t=1$ corresponding to the dynamics of the system when three different ensembles of initial conditions for the position are considered:  steady state, Gaussian distributed and a delta function at zero position. The inset shows the ratio $\ln p(R)/p(-R)$ as a function of $R$ for all three initial distributions. One can clearly see that only for the steady state initial ensemble the detailed fluctuation theorem holds. It is worthy to stress this point to prevent confusions and errors when applying  the predictions provided by the theorem in real experiments, in which one should have careful control on the preparation of the initial state of the system. It would be interesting, although more tedious, to test the predictions derived from the fluctuation theorem away from the steady state, in which no simple expression such as Eq.~(\ref{Rmain}) is available.

\subsection{Mean dissipated work and $\langle R \rangle$} 

For dissipative systems in a steady state, $R$ is closely related to the so called housekeeping heat $Q_{hk}$ . This is the heat that is permanently dissipated while maintaining a nonequilibrium state at fixed external parameters. It has been shown \cite{hh} that  $R=Q_{hk}/T$, at the same time that
\begin{equation} \label{RW}
\langle Q_{hk} \rangle =   t J^{2}  \int_{0}^{L} dx \frac{1}{p(x)}. 
\end{equation}
The steady flux $J$ is related to the probability $p(x)$ according to Eq.~(\ref{eq:flux_diff}), which makes it possible to express the housekeeping heat in terms of the external driving and the particle flux as
$\langle Q_{hk} \rangle =  t  F  J L $. Since  $J L = \langle \dot{x} \rangle$,  the mean heat produced can be rewritten in the suggestive form
\begin{equation}
\langle Q_{hk} \rangle =  F \langle \dot{x} \rangle   t,
\end{equation}
which implies that the mean total entropy change can be simply expressed  in terms of the mean work produced and dissipated by the driving force,  $\langle W \rangle =F \langle  \Delta x \rangle$, as
\begin{equation}
\langle R \rangle = \langle W \rangle / T.
\end{equation}
We can derive an explicit expression for the mean velocity of the particle~ \cite{reim}
\begin{equation}
\langle \dot{x} \rangle = \frac{ L T (1-e^{-2 \pi f})  }
  { \int_{0}^{L} dxe^{-U(x)/T}  \int_{x}^{x + L} dy e^{U(y)/T}
  }.
\end{equation}
Remember that $U(x)=V(x)-Fx$. Again, for  the potential $V(x)$ given by Eq.~(\ref{pot})  with $v_{0}=1$,  one gets, after some integrations,
\begin{equation}
\langle \dot{x} \rangle = \frac{2 \pi T}{L} \frac{f^{3}+f}{f^{2}+2/ \sqrt{3}}.
\end{equation}
Note that the mean velocity could be recasted 
in terms of an effective mobility $\mu_{eff}$, which would depend on the applied force, so that 
$\langle \dot{x} \rangle = F  \mu_{eff}$,
For small forces, $\mu_{eff}= \sqrt{3}/2$ which would correspond to the free diffusion of a Brownian particle in the periodic potential, and for asymptotically large forces, $\mu_{eff}$ reduces to the bare mobility, $\mu_{eff}=1$, in scaled units.

\begin{figure} 
\begin{center}
  \includegraphics[ angle=270, width=9cm]{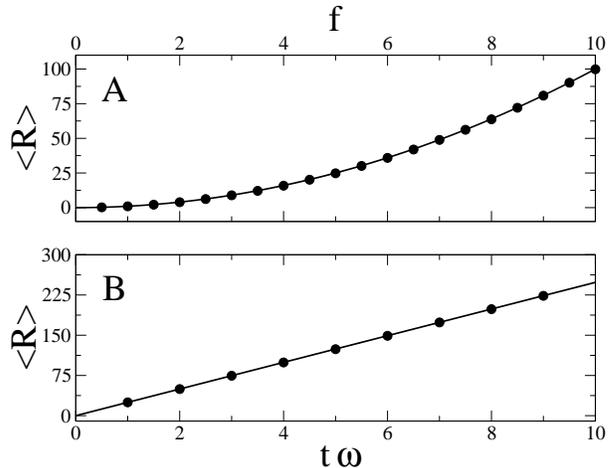}
  \caption{Plot of $\langle R \rangle$ versus the external force $f$ for $t \omega=1$
    (A) and versus the dimensionless time $t \omega$ for $f=5$ (B). Dots are obtained from numerical simulations, while solid lines
    correspond to the exact analytical expression  (\ref{meanRexact}).}
  \label{meanR}
\end{center}
\end{figure}

Finally, coming back to the mean entropy production, the following analytical expression is then encountered,
\begin{equation} \label{meanRexact}
\langle R \rangle = t \omega f^{2} \frac{f^{2}+1}{f^{2}+2/ \sqrt{3}},
\end{equation}
where $\omega=T (2 \pi / L)^2$.
 In Fig. (\ref{meanR}) we compare the theoretical prediction for  $\langle R \rangle$ with numerical results, exploring the force $f$ and the time $t$. Theory and simulations are in excellent agreement. The linear dependence in time of the mean value of $R$ yields to a constant mean entropy production rate.

\subsection{"Transient-violation" trajectories}

\begin{figure} 
\begin{center}
  \includegraphics[ angle=270, width=9cm]{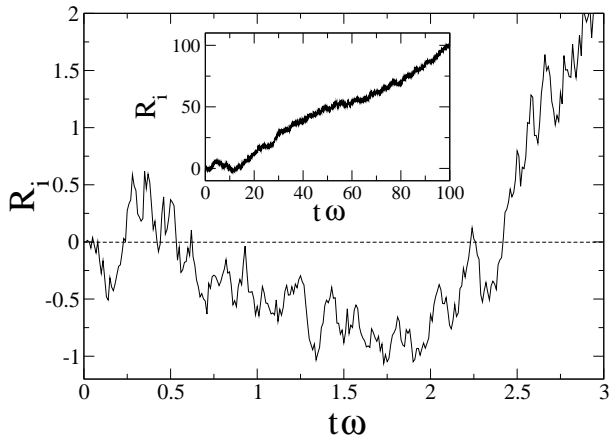}
  \caption{Total entropy change along one single trajectory $R_i$ as a function
    of time for $f=1$. Note that, for short times, "transient
    violations of the second law" are observable since the total $R$  is
    negative. In the inset the behavior of $R$ at larger times is plotted, where
    fluctuations are less dominant. Averaging over a large number of independent trajectories would lead to the linear increase of   $\langle R \rangle$ predicted by Eq. (\ref{meanRexact}). }
  \label{transient}
\end{center}
\end{figure}

Although $\langle R \rangle$ increases linearly in time, for a single trajectory $R$ will fluctuate. As a result, for short times and for typical energies of the system comparable to the thermal energy, negative values of $R$ will be observable, corresponding to the so-called "transient violations of the second law".  These violations contribute significantly to the averages at the same time that determine the tails of the distributions, which are crucial when examining the detailed fluctuation relation.
In fact, negative values of $R$ are directly related to Figs. (\ref{e-R}), (\ref{p(R)}) and (\ref{StrongRel}), and their corresponding subsections.

In Fig. (\ref{transient}) we plot the time evolution of the entropy production $R$ of a single trajectory. One can see that at short times $R$ is likely to become negative, indicating that up to that point the total entropy  production associated to that stochastic trajectory is negative. The inset shows that at long times $R$ is positive and increases. Upon averaging over a number of independent trajectories, the linear increase predicted in the previous section is recovered. Hence, the study of single trajectories highlights the stochastic nature of fluctuation theorems and the need to distinguish between mean response of the system and the properties of individual trajectories. As the period of observation increases, it becomes rapidly unlikely that negative values of $R$ are observed.  This, together with the energy scale, sets  the regime where examining  "transient violations" becomes impractical.
 
\section{Conclusions and final remarks}

We have considered  a simple system in a nonequilibrium steady state and have studied in detail many of the predictions and consequences derived from the fluctuation theorem in Ref. \cite{sei}. In
particular, by choosing an appropriate periodic potential, we have been able to find an explicit analytical and physically meaningful expression for the total entropy production along a stochastic trajectory.
Advantage has been taken from the fact that  only initial and final positions for each trajectory 
are relevant, being a potential useful feature in future experiments.
Numerical
simulations of the dynamics of the Langevin equation have been necessary to build the
probability distribution functions of $R$. 
We have elucidated the relevance and properties of $p(R)$ in the system. 
Moreover, all of the results based, related, derived and predicted from the fluctuation relation have
been tested satisfactorily.

This work leads to a more profound understanding of the specific
meaning and practical appliance of such abstract general results. At the same time, it opens the door to study new and more complex cases such as interacting particles or the effect of temperature dependent substrates.  Furthermore, it would be very interesting to analyze fluctuations observed in biological systems such as  molecular motors under the perspective of fluctuation relations.
Studies based on simple physical models are crucial
to illustrate the key ingredients and 
foresee the consequences of these theoretical results on
real experiments. Therefore, they bridge the 
fundamental advances in statistical mechanics with the active area of nano 
scale experiments on biological systems.

We thank Udo Seifert for useful discussions. 
This research was supported by the Ministerio de Educaci\'on y Ciencia (Spain)
under the grant  FPU-AP-2004-0770 (A. G-M.) and by DGICYT of the Spanish Government and Distinci\'o de la Generalitat de Catalunya (Spain) (I. P.).

\end{document}